\documentclass[10pt,draftcls,journal,letterpaper,twoside,onecolumn]{IEEEtran}
\usepackage{hyperref}
\usepackage{ifpdf}
\usepackage{ifthen,calc}
\usepackage{graphicx}
\usepackage{psfrag}
\usepackage[cmex10]{amsmath}
\usepackage{amssymb,amsfonts,eucal}
\usepackage{cite}
\usepackage{color}
\usepackage{ulem}
\ifpdf
 \DeclareGraphicsExtensions{.pdf,.png,.jpg,.mps}
\else \DeclareGraphicsExtensions{.eps} \fi
\hypersetup{
    pdftitle={Single pump, parametric amplification in randomly-birefringent,
unidirectionally spun fibers},    
    pdfauthor={Marco Santagiustina, Luca Schenato},     
    pdfsubject={paper for IEEE PHOTONICS TECHNOLOGY LETTERS},   
    pdfkeywords={Optical fiber amplifiers, Optical fiber dispersion, Optical fiber
polarization, Spun fibers.} 
}
\begin{document}
\newcommand{\ket}[1]{|#1\rangle}
\newcommand{\outbracket}[2]{\langle #1 | #2 \rangle}
\newcommand{\inbracket}[2]{| #1 \rangle \langle #2|}
\newcommand{\abs}[1]{\left| #1 \right|}
\newcommand{\mean}[1]{\langle #1 \rangle}
\newcommand{\stokes}[3]{\langle #1 | #2 | #3 \rangle}
\title{Single pump, parametric amplification in randomly-birefringent,
unidirectionally spun fibers}
\author{
M.~Santagiustina,~\IEEEmembership{Member,~IEEE,}
L.~Schenato,~\IEEEmembership{Member,~IEEE.}
\thanks{A.~Santagiustina, L.~Schenato are with the
   Department of Information Engineering, University of Padova,
   Via G.~Gradenigo 6/B, 35131 Padova, Italy,
   e-mail: name.surname@dei.unipd.it.}%
   \thanks{The research leading to these results has received funding from the European
Community's Seventh Framework Programme under grant agreement n 219299, Gospel.
The research was also held in the framework of the agreement with ISCOM (Rome).}
\thanks{Digital Object Identifier 00.0000/LPT.2009.000000}}
\markboth{IEEE PHOTONICS TECHNOLOGY LETTERS,~VOL.~XX,~NO.~YY,~MONTH~YEAR}
   {SANTAGIUSTINA \MakeLowercase{\textit{et al.}}: Single pump, parametric....}
%
%
\maketitle
%
%
%
%
\begin{abstract}
A systematic study of the effects of polarization mode dispersion on
broad-band and narrow-band, single pump, fiber parametric amplifiers
is realized through numerical solutions of the equations governing the
interaction. The nonlinear polarization rotation is shown to be a relevant effect
that can increase gain randomness when it mixes with polarization
mode dispersion. In unidirectionally spun fibers
the signal-pump alignment can be highly increased and the gain enhanced.
However, in spite of the enhanced alignment, large polarization mode dispersion phase-mismatches the interaction and the gain decreases to zero.
\end{abstract}
%
%
%
%
%
\section{Introduction}
Fiber optical parametric amplifiers (FOPAs) have been under investigation for many
years. In the single pump configuration both broad-band (BB-FOPA) \cite{mar96ol} and narrow-band
(NB-FOPA) \cite{HAR03OL,MAR04JSQE} amplification can be obtained.
The former has been mainly addressed for amplification and wavelength conversion
\cite{agrabook} while the latter yields a superb technique for slow and fast light 
generation \cite{DAH05OE,sch08jlt}.

FOPAs are highly affected by the fiber random birefringence (polarization mode dispersion - PMD) 
\cite{lin04ol,wil08jlt}.
In an ideal, perfectly isotropic fiber the waves would maintain their input states 
of polarization (SOP) over the entire fiber, the gain would be deterministic and 
it would depend only on the initial SOPs \cite{mar06jlt}.
In low birefringence fibers the evolution of the pump, signal and idler SOPs
is random, due to the PMD. Eventually, the gain is reduced and becomes
a random quantity, as theoretically analyzed in \cite{lin04ol} for BB-FOPA.

Recently, the benefits of unidirectional fiber spinning in Raman \cite{bet08ptl}
and Brillouin fiber amplifiers \cite{gal08ptl} have been underlined. 
Here, the same detailed study is extended to single pump, BB- and NB-FOPAs. 
Note that, differently from the Raman and Brillouin fiber amplifiers,
in FOPAs pump polarization scrambling cannot be used to avoid
polarization dependence. Other techniques like dual pumping \cite{INO92JQE,jop93el} or polarization diversity
\cite{HAS93PTL,WON02PTL} were proposed to overcome the problem, but at the cost of an increased 
set-up complexity.
So, the results presented here are particularly relevant because
unidirectionally spun fibers can represent a breakthrough for implementing
highly performant BB- and NB-FOPA with simple experimental setups.
\section{Propagation model}
The parametric interaction of the pump, signal and idler
monochromatic complex Jones vectors, $\ket{A_p}, \ket{A_s}, \ket{A_i}$,
in the undepleted pump approximation and neglecting
the nonlinear effects of the signal and idler on the pump, is governed by \cite{lin04ol}:
\begin{eqnarray}
\label{fopaeq}
&\partial_z \ket{A_p} =j \left( j \alpha_p + k(\omega_p)- \bar{\beta}(\omega_p) \cdot \bar{\sigma} /2
\right) + \nonumber \\
&+ j \gamma/3 \left( 2 \outbracket{A_p}{A_p} + \inbracket{A_p^*}{A_p^*} \right) \ket{A_p}, \nonumber \\
&\partial_z \ket{A_{s,i}} = j \left( j \alpha_{s,i} + k(\omega_{s,i})- \bar{\beta}(\omega_{s,i})
\cdot \bar{\sigma} /2 \right) \ket{A_{s,i}} + \nonumber \\
&+ j 2 \gamma/3 \left( 2 \outbracket{A_p}{A_p} + \inbracket{A_p}{A_p} + \inbracket{A_p^*}{A_p^*} \right)
\ket{A_{s,i}} + \nonumber \\
& + j \gamma/3 \left( \outbracket{A_p^*}{A_p} + 2 \inbracket{A_p}{A_p^*} \right)
\ket{A_{i,s}^*},
\end{eqnarray}
where $\gamma=2 \; W^{-1} km^{-1}$ is the nonlinear coefficient,
$\alpha_h, \; h=p,s,i$ are the loss coefficients
at the optical pulsations of the pump, signal and idler $\omega_p, \omega_s, \omega_i$ 
that satisfy simultaneously the energy $2\omega_p=\omega_s+\omega_i$ and 
the nonlinear phase matching condition
$\Delta k= k(\omega_s)+k(\omega_i)-2 k(\omega_p) \approx k_{2p}\, (\omega_s-\omega_p)^2+k_{4p}\,(\omega_s-\omega_p)^4/12= -2 \gamma P_0$,
($k_{nh}$ is the $n$-th angular frequency derivative of
$k(\omega)$, evaluated at $\omega_h$ \cite{mar96ol} and
$P_0$ is the pump power).
Transition from BB- to NB-FOPA can be obtained by tuning $\omega_p$ in the vicinity of 
the fiber zero dispersion wavelength, thus modifying the dispersion coefficient $k_{2p}$.
When the sign of $k_{2p}$ is negative, gain occurs at $\omega_s$ within a broad band around the pump frequency. 
For positive $k_{2p}$ and if $k_{4p}$ is negative, the nonlinear matching can be satisfied 
in two narrow spectral bands, largely detuned from the pump pulsation.
The following parameters have been used in the simulations:
BB-FOPA, $k_{2p}=-1.18\times 10^{-29}\;\mathrm{s}^2/\mathrm{m}$, $k_{4p}=1\times 10^{-55}\;\mathrm{s}^4/\mathrm{m}$ \cite{lin04ol};
NB-FOPA, $k_{2p}=7.68\times 10^{-28}\;\mathrm{s}^2/\mathrm{m}$, $k_{4p}=-5\times 10^{-55}\;\mathrm{s}^4/\mathrm{m}$ \cite{MAR04JSQE}.

The Stokes vector $\bar{\beta}(z, \omega)$ is the local birefringence vector
and $\bar{\sigma}$ is the vector of the Pauli spin matrices \cite{gor00pnas}.
For unspun fibers, $\bar{\beta}(z,\omega)=\bar{\beta}_{us} (z,\omega)$ 
can be obtained through the so called random modulus model (RMM)
\cite{wai97josab}, that describes the PMD phenomenon and effects
through three characteristic lengths: the beat length ($L_B$),
the birefringence correlation length ($L_F$) and the polarization correlation length ($L_C$). 
The first depends on the pulsation through $L_B(\omega)=\omega_0 L_B(\omega_0)/\omega$; the 
latter is independent of $\omega$ ($L_F=9~\mathrm{m}$ was fixed in the numerical integrations).
The first two contribute to determine the PMD coefficient (PMDC),
hereinafter defined as $D=\sqrt{\langle \Delta \tau^2 \rangle/L}$ where
$\langle \Delta \tau^2 \rangle$ is the fiber PMD mean 
square differential group delay and $L(=2~\mathrm{km})$ is the fiber length.
The polarization correlation length $L_C$, is the length scale over which the SOP
is modified and depends on the previously defined lengths and on the fiber spinning properties.
This length scale is a fundamental parameter when nonlinear PMD sets in 
\cite{wai97josab}, as in this case.

For unidirectionally spun fibers the birefringence vector becomes 
$\bar{\beta}(z,\omega) = R_3[2\phi(z)] \bar\beta_{us} (z,\omega)$
where $R_{3}$ is a Mueller matrix representing a rotation around 
the third component of the Stokes space ($\hat u_3$). 
The angle of rotation is given by the spin function $\phi(z)=2\pi z/p$ 
where $p$ is called the spin pitch.
It was found that unidirectional spinning \cite{gal06oft}: a) decreases the PMDC and b) modifies the 
polarization correlation length $L_C$, according to:
\begin{eqnarray}
\label{lcorr}
L_C=\frac{L_B^2}{a \pi^2 L_F} \left(1+ \frac{16 \pi^2 L_F^2}{p^2} \right)
\end{eqnarray}
for $L_B^2 >> 4 \pi^2 L_F^2 / \sqrt{1+(16 \pi^2 L_F^2)/p^2}$ and where $a=4$ ($a=2$)
for linear (circular) input SOPs.

A few thousands statistical realizations of the stochastic processes, using the RMM,
and subsequent integrations of Eqs. \ref{fopaeq}, have been performed to determine the 
mean gain $\mu_G=\mean{\outbracket{A_s(L)}{A_s^*(L)}}/ \outbracket{A_s(0)}{A_s^*(0)}$ 
and the relative gain standard deviation (STD) 
$\sigma_G=\sqrt{ \mean{ \outbracket{A_s(L)}{A_s^*(L)} } / \mean{\sqrt{\outbracket{A_s(L)}{A_s^*(L)}}}^2 -1}$.
Numerical results are represented in all figures as functions of the PMDC
that is calculated from the statistical ensemble.
\section{Numerical results}
\begin{figure}[htb]
%
%
\begin{psfrags}%
\psfragscanon%
%
\psfrag{s03}[t][b]{\color[rgb]{0,0,0}\setlength{\tabcolsep}{0pt}\begin{tabular}{c}$\mu_G [\mathrm{dB}]$\end{tabular}}%
\psfrag{s06}[b][b]{\color[rgb]{0,0,0}\setlength{\tabcolsep}{0pt}\begin{tabular}{c}$D [\mathrm{ps} \times \mathrm{km}^{-1/2}]$\end{tabular}}%
\psfrag{s07}[t][b]{\color[rgb]{0,0,0}\setlength{\tabcolsep}{0pt}\begin{tabular}{c}$\sigma_G [\mathrm{dB}]$\end{tabular}}%
%
\psfrag{x01}[t][t]{0}%
\psfrag{x02}[t][t]{0.1}%
\psfrag{x03}[t][t]{0.2}%
\psfrag{x04}[t][t]{0.3}%
\psfrag{x05}[t][t]{0.4}%
\psfrag{x06}[t][t]{0.5}%
\psfrag{x07}[t][t]{0.6}%
\psfrag{x08}[t][t]{0.7}%
\psfrag{x09}[t][t]{0.8}%
\psfrag{x10}[t][t]{0.9}%
\psfrag{x11}[t][t]{1}%
\psfrag{x12}[t][t]{$10^{-4}$}%
\psfrag{x13}[t][t]{$10^{-3}$}%
\psfrag{x14}[t][t]{$10^{-2}$}%
\psfrag{x15}[t][t]{$10^{-1}$}%
\psfrag{x16}[t][t]{$10^{0}$}%
\psfrag{x17}[t][t]{}%
\psfrag{x18}[t][t]{}%
\psfrag{x19}[t][t]{}%
\psfrag{x20}[t][t]{}%
\psfrag{x21}[t][t]{}%
%
\psfrag{v01}[r][r]{0}%
\psfrag{v02}[r][r]{0.1}%
\psfrag{v03}[r][r]{0.2}%
\psfrag{v04}[r][r]{0.3}%
\psfrag{v05}[r][r]{0.4}%
\psfrag{v06}[r][r]{0.5}%
\psfrag{v07}[r][r]{0.6}%
\psfrag{v08}[r][r]{0.7}%
\psfrag{v09}[r][r]{0.8}%
\psfrag{v10}[r][r]{0.9}%
\psfrag{v11}[r][r]{1}%
\psfrag{v12}[r][r]{0}%
\psfrag{v13}[r][r]{}%
\psfrag{v14}[r][r]{2}%
\psfrag{v15}[r][r]{}%
\psfrag{v16}[r][r]{4}%
\psfrag{v17}[r][r]{}%
\psfrag{v18}[r][r]{0}%
\psfrag{v19}[r][r]{10}%
\psfrag{v20}[r][r]{20}%
\psfrag{v21}[r][r]{30}%
%
\includegraphics[width=\columnwidth]{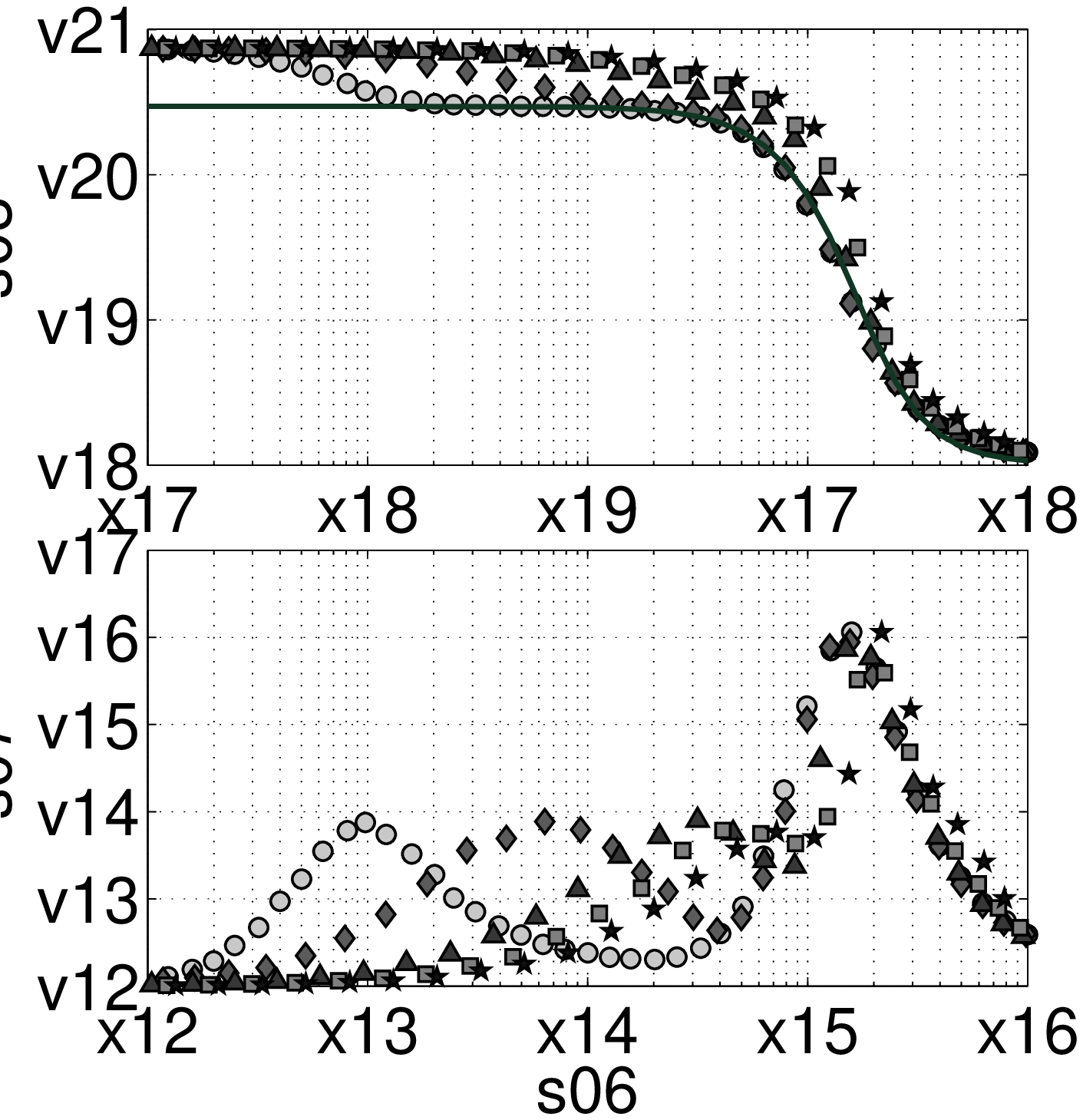}%
\end{psfrags}%
%

\caption{Mean gain and STD for initially linear and parallel SOPs for BB-FOPA as a function of $D$. 
For unspun fibers, the solid curve is the result of the mean equations of \cite{lin04ol},
while the circles are the result of the numerical integration of Eqs. \ref{fopaeq}.  
The numerical results for spun fibers with spin rate $\nu=1/p=0.5,\,1,\,2$ and $10$ turns/m 
are represented respectively by diamonds, triangles squares and stars.}
\label{fig:figura1}
\end{figure}

\begin{figure}[htb]
%
%
\begin{psfrags}%
\psfragscanon%
%
\psfrag{s03}[t][b]{\color[rgb]{0,0,0}\setlength{\tabcolsep}{0pt}\begin{tabular}{c}$\mu_G~[\mathrm{dB}]$\end{tabular}}%
\psfrag{s06}[b][b]{\color[rgb]{0,0,0}\setlength{\tabcolsep}{0pt}\begin{tabular}{c}$D [\mathrm{ps} \times \mathrm{km}^{-1/2}]$\end{tabular}}%
\psfrag{s07}[t][b]{\color[rgb]{0,0,0}\setlength{\tabcolsep}{0pt}\begin{tabular}{c}$\sigma_G [\mathrm{dB}]$\end{tabular}}%
%
\psfrag{x01}[t][t]{0}%
\psfrag{x02}[t][t]{0.1}%
\psfrag{x03}[t][t]{0.2}%
\psfrag{x04}[t][t]{0.3}%
\psfrag{x05}[t][t]{0.4}%
\psfrag{x06}[t][t]{0.5}%
\psfrag{x07}[t][t]{0.6}%
\psfrag{x08}[t][t]{0.7}%
\psfrag{x09}[t][t]{0.8}%
\psfrag{x10}[t][t]{0.9}%
\psfrag{x11}[t][t]{1}%
\psfrag{x12}[t][t]{$10^{-4}$}%
\psfrag{x13}[t][t]{$10^{-3}$}%
\psfrag{x14}[t][t]{$10^{-2}$}%
\psfrag{x15}[t][t]{$10^{-1}$}%
\psfrag{x16}[t][t]{$10^{0}$}%
\psfrag{x17}[t][t]{}%
\psfrag{x18}[t][t]{}%
\psfrag{x19}[t][t]{}%
%
\psfrag{v01}[r][r]{0}%
\psfrag{v02}[r][r]{0.1}%
\psfrag{v03}[r][r]{0.2}%
\psfrag{v04}[r][r]{0.3}%
\psfrag{v05}[r][r]{0.4}%
\psfrag{v06}[r][r]{0.5}%
\psfrag{v07}[r][r]{0.6}%
\psfrag{v08}[r][r]{0.7}%
\psfrag{v09}[r][r]{0.8}%
\psfrag{v10}[r][r]{0.9}%
\psfrag{v11}[r][r]{1}%
\psfrag{v12}[r][r]{0}%
\psfrag{v13}[r][r]{}%
\psfrag{v14}[r][r]{2}%
\psfrag{v15}[r][r]{}%
\psfrag{v16}[r][r]{4}%
\psfrag{v17}[r][r]{}%
\psfrag{v18}[r][r]{0}%
\psfrag{v19}[r][r]{10}%
\psfrag{v20}[r][r]{20}%
\psfrag{v21}[r][r]{30}%
%
\includegraphics[width=\columnwidth]{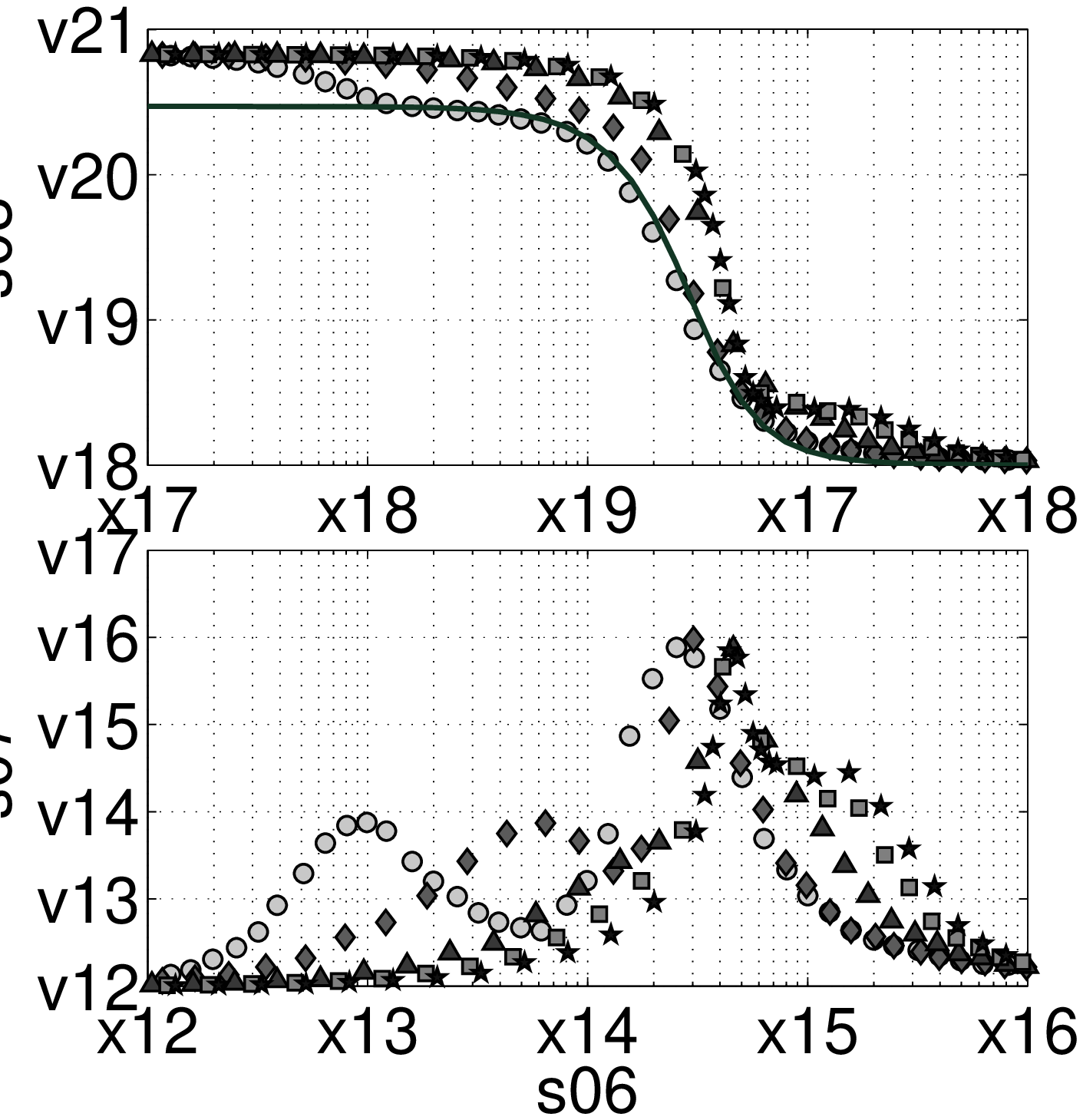}%
\end{psfrags}%
%

\caption{Mean gain and STD for initially linear and parallel SOPs for NB-FOPA as a function of $D$. 
Symbols are the same of Fig. \ref{fig:figura1}.}\label{fig:figura2}
\end{figure}

For unspun fibers, the mean gain and STD from numerical solutions are presented (circles)
for initially linear and parallel SOPs in Fig.~\ref{fig:figura1} for BB-FOPAs \uline{($\lambda_p\simeq 1550.2$~nm, $\lambda_s\simeq 1514$~nm)}  and in Fig.~\ref{fig:figura2}
for NB-FOPA \uline{($\lambda_p\simeq 1541$~nm, $\lambda_s\simeq 1387$~nm)} as a function of $D$. Mean and STD deviation for other
input SOPs are not presented here, for the sake of brevity, but all the results show
the same behavior that is described below.
When the PMDC is large, the gain decreases down to zero, because
PMD tends to phase mismatch the waves \cite{lin04ol,wil08jlt}. In this regime
the predictions of the mean Eqs. of \cite{lin04ol} (solid curves in both figures),
where nonlinearity was averaged and an effective nonlinear coefficient $\gamma_e=8\gamma/9$ was found \cite{WAI91OL,Evangelides1992}, 
are recovered both for BB- and NB-FOPA.
The PMD affects more NB-FOPA than BB-FOPA mean gain; this can be explained by
the fact that the effects of PMD, at the same value of $D$, increases with
the frequency detuning among the waves.


\begin{figure}[htb]
%
%
\begin{psfrags}%
\psfragscanon%
%
\psfrag{s06}[b][b]{\color[rgb]{0,0,0}\setlength{\tabcolsep}{0pt}\begin{tabular}{c}$p~[\mathrm{m}]$\end{tabular}}%
\psfrag{s01}[b][b]{\color[rgb]{0,0,0}\setlength{\tabcolsep}{0pt}\begin{tabular}{c}$\nu~[\mathrm{turns/m}]$\end{tabular}}%
\psfrag{s02}[b][b]{\color[rgb]{0,0,0}\setlength{\tabcolsep}{0pt}\begin{tabular}{c}$L_B~[\mathrm{m}]$\end{tabular}}%
%

\psfrag{x12}[t][t]{0}%
\psfrag{x13}[t][t]{}%
\psfrag{x14}[t][t]{0.2}%
\psfrag{x15}[t][t]{}%
\psfrag{x16}[t][t]{0.4}%
\psfrag{x17}[t][t]{}%
\psfrag{x18}[t][t]{0.6}%
\psfrag{x19}[t][t]{}%
\psfrag{x20}[t][t]{0.8}%
\psfrag{x21}[t][t]{}%
\psfrag{x22}[t][t]{1}%

\psfrag{x23}[B][B]{Inf}%
\psfrag{x24}[B][B]{10}%
\psfrag{x25}[B][B]{}%
\psfrag{x26}[B][B]{3.33}%
\psfrag{x27}[B][B]{}%
\psfrag{x28}[B][B]{2.0}%
\psfrag{x29}[B][B]{}%
\psfrag{x30}[B][B]{1.43}%
\psfrag{x31}[B][B]{}%
\psfrag{x32}[B][B]{1.11}%
\psfrag{x33}[B][B]{1}%
%
\psfrag{v01}[r][r]{0}%
\psfrag{v02}[r][r]{0.1}%
\psfrag{v03}[r][r]{0.2}%
\psfrag{v04}[r][r]{0.3}%
\psfrag{v05}[r][r]{0.4}%
\psfrag{v06}[r][r]{0.5}%
\psfrag{v07}[r][r]{0.6}%
\psfrag{v08}[r][r]{0.7}%
\psfrag{v09}[r][r]{0.8}%
\psfrag{v10}[r][r]{0.9}%
\psfrag{v11}[r][r]{1}%
\psfrag{v12}[r][r]{$10^{0}$}%
\psfrag{v13}[r][r]{$10^{1}$}%
\psfrag{v14}[r][r]{$10^{2}$}%
\psfrag{v15}[r][r]{}%
%
\includegraphics[width=\columnwidth]{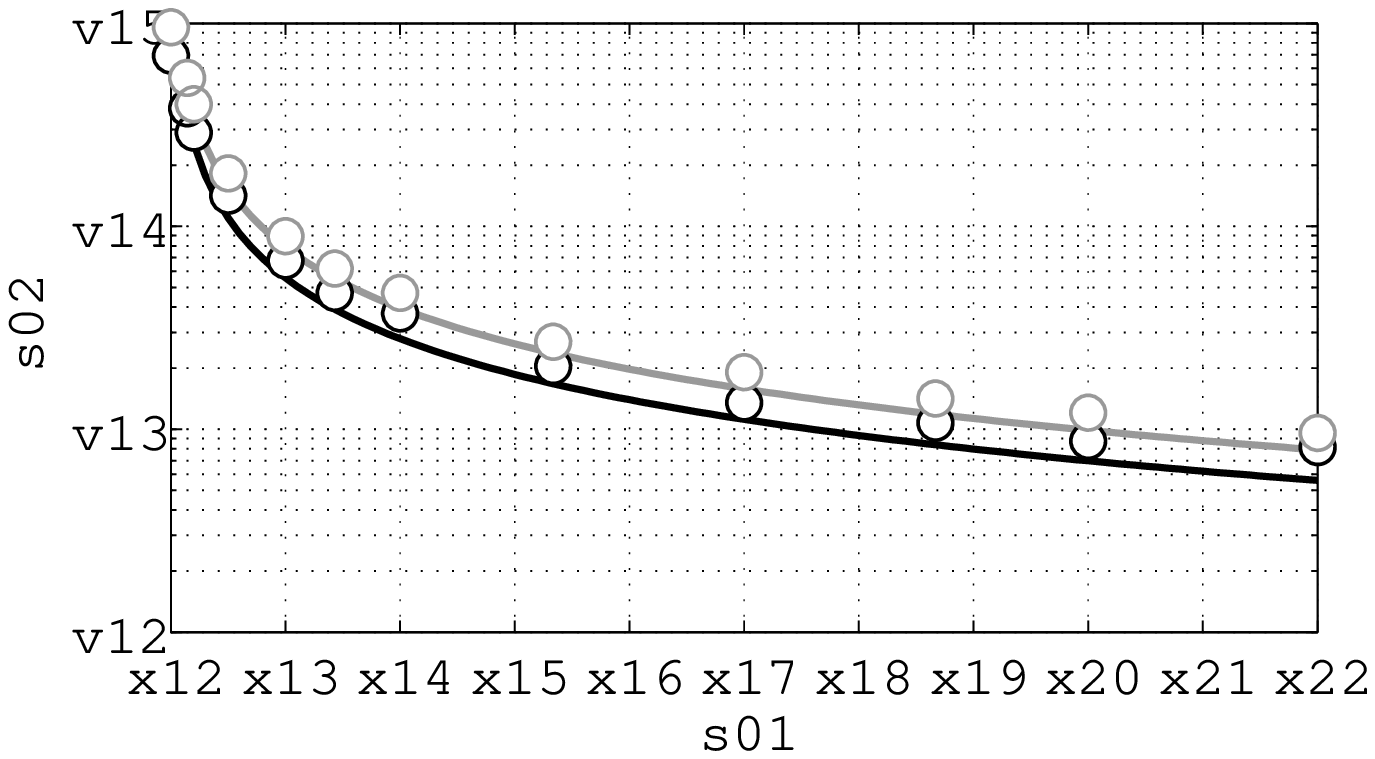}%
\end{psfrags}%
%

\caption{Values of $L_B$ vs. $\nu=1/p$ satisfying the condition 
$L_C = L_{NPR}$ in BB-FOPA: theoretical (solid curves) and numerical (circles) solutions are represented for both linear (black) and circular (gray) input SOPs.}\label{fig:figura3}
\end{figure}

However, for small PMDC the mean and STD values do not correspond to the theoretical findings.
The mean gain is actually larger and the STD presents a peak
(see $\sigma_G$ in Figs. \ref{fig:figura1} and \ref{fig:figura2} for $D \approx 10^{-3}\;\mathrm{ps}/\sqrt{\mathrm{km}}$)
not predicted by the theory.
The discrepancy stems from the nonlinear PMD \cite{wai97josab}, induced by the
nonlinear polarization rotation (NPR) effect, as already briefly described
in the case of Raman amplification \cite{gal06jlt}.
In the Stokes space, the pump vector $\bar{P}=\stokes{A_p}{\bar{\sigma}}{A_p}$, is governed by:
\begin{equation}
\label{pump}
\partial_z \bar{P} =[\bar{\beta}(z,\omega_p) + 2 \gamma /3 (\bar{P} \cdot \hat u_3) \hat u_3] \times \bar{P}.
\end{equation}
When the pump polarization evolution length $L_C$ is close to the NPR length scale
NPR becomes a significant source of randomness \cite{wai97josab}, which
finally leads to enhance signal gain randomness.
To prove this affirmation, the theoretical values of $L_B$ satisfying the condition 
$L_C = L_{NPR}$ in Eq. \ref{lcorr}, in function of the spin rate $\nu=1/p$, 
have been calculated for the BB-FOPA and are shown in Fig. \ref{fig:figura3} (solid curves),
in the limit of validity of Eq. \ref{lcorr} (similar results are
obtained for the NB-FOPA).
The value of the NPR length was estimated as $L_{NPR} \approx 9 /(2 \gamma P_0)=2.25$ km
in the hypothesis, which is valid if the nonlinear mixing is strong 
enough, that the mean power in the stochastic vector $(\bar{P} \cdot \hat u_3) \hat u_3$ 
is one third of the total pump power.
In the same Fig. \ref{fig:figura3} the values of $L_B$ for which the STD reaches the relative maximum
in the numerical solutions are also shown for comparison (circles). 
The agreement is very good both for linear (black solid curve and circles) 
and circular (grey solid curve and circles) input SOPs.

\begin{figure}[htb]
%
%
\begin{psfrags}%
\psfragscanon%
%
\psfrag{s02}[b][b]{\color[rgb]{0,0,0}\setlength{\tabcolsep}{0pt}\begin{tabular}{c}$z [\mathrm{km}]$\end{tabular}}%
\psfrag{s03}[t][b]{\color[rgb]{0,0,0}\setlength{\tabcolsep}{0pt}\begin{tabular}{c}$\mean{cos(\theta)}$\end{tabular}}%
\psfrag{s09}[l][lb]{\color[rgb]{0,0,0}\setlength{\tabcolsep}{0pt}\begin{tabular}{l}\end{tabular}}%
\psfrag{s10}[l][lb]{\color[rgb]{0,0,0}\setlength{\tabcolsep}{0pt}\begin{tabular}{l}\end{tabular}}%
\psfrag{s11}[l][lb]{\color[rgb]{0,0,0}\setlength{\tabcolsep}{0pt}\begin{tabular}{l}\end{tabular}}%

\psfrag{s13}[l][lb]{\color[rgb]{0,0,0}\setlength{\tabcolsep}{0pt}\begin{tabular}{l}\end{tabular}}%
%
\psfrag{x01}[t][t]{0}%
\psfrag{x02}[t][t]{0.1}%
\psfrag{x03}[t][t]{0.2}%
\psfrag{x04}[t][t]{0.3}%
\psfrag{x05}[t][t]{0.4}%
\psfrag{x06}[t][t]{0.5}%
\psfrag{x07}[t][t]{0.6}%
\psfrag{x08}[t][t]{0.7}%
\psfrag{x09}[t][t]{0.8}%
\psfrag{x10}[t][t]{0.9}%
\psfrag{x11}[t][t]{1}%
\psfrag{x12}[t][t]{0}%
\psfrag{x13}[t][t]{}%
\psfrag{x14}[t][t]{0.5}%
\psfrag{x15}[t][t]{}%
\psfrag{x16}[t][t]{1.0}%
\psfrag{x17}[t][t]{}%
\psfrag{x18}[t][t]{1.5}%
\psfrag{x19}[t][t]{}%
\psfrag{x20}[t][t]{2.0}%
%
\psfrag{v01}[r][r]{0}%
\psfrag{v02}[r][r]{0.1}%
\psfrag{v03}[r][r]{0.2}%
\psfrag{v04}[r][r]{0.3}%
\psfrag{v05}[r][r]{0.4}%
\psfrag{v06}[r][r]{0.5}%
\psfrag{v07}[r][r]{0.6}%
\psfrag{v08}[r][r]{0.7}%
\psfrag{v09}[r][r]{0.8}%
\psfrag{v10}[r][r]{0.9}%
\psfrag{v11}[r][r]{1}%
\psfrag{v12}[r][r]{0.3}%
\psfrag{v13}[r][r]{}%
\psfrag{v14}[r][r]{0.4}%
\psfrag{v15}[r][r]{}%
\psfrag{v16}[r][r]{0.5}%
\psfrag{v17}[r][r]{}%
\psfrag{v18}[r][r]{0.6}%
\psfrag{v19}[r][r]{}%
\psfrag{v20}[r][r]{0.7}%
\psfrag{v21}[r][r]{}%
\psfrag{v22}[r][r]{0.8}%
\psfrag{v23}[r][r]{}%
\psfrag{v24}[r][r]{0.9}%
\psfrag{v25}[r][r]{}%
\psfrag{v26}[r][r]{1}%
%
\includegraphics[width=\columnwidth]{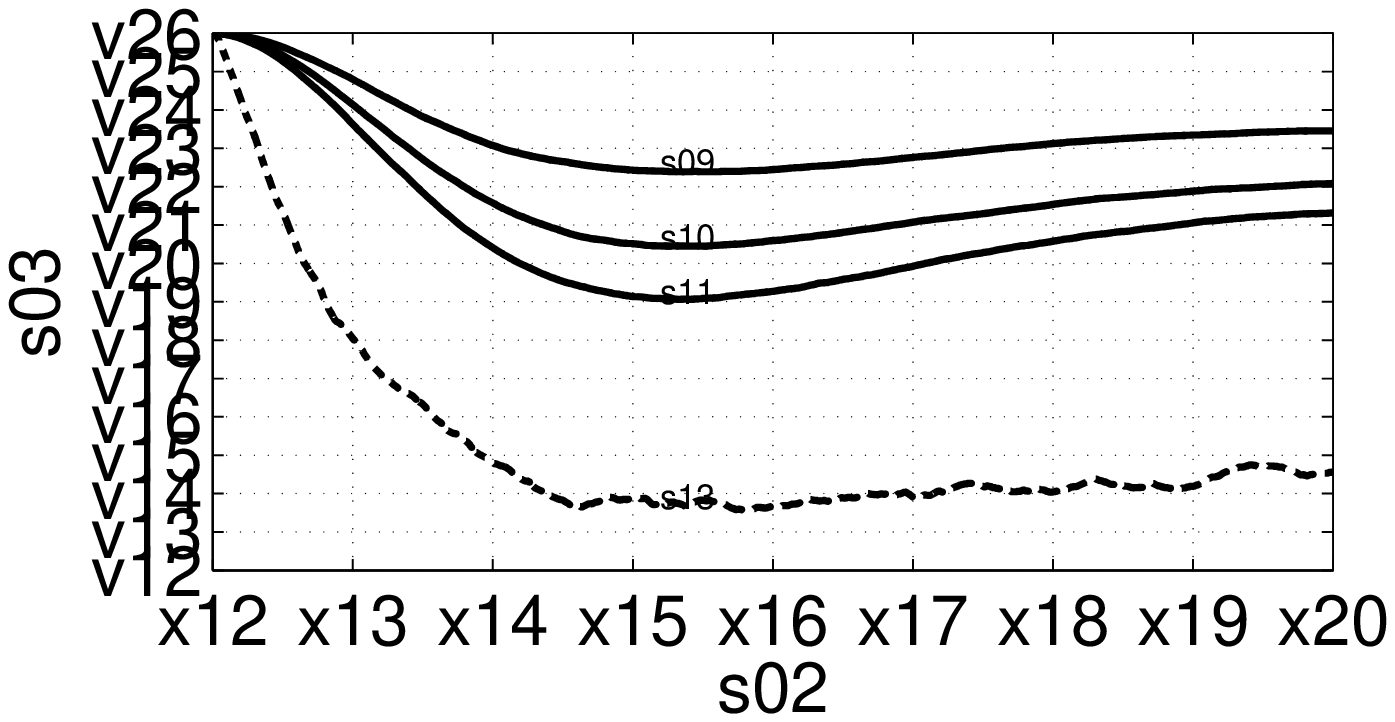}%
\end{psfrags}%
%

\caption{Evolution of $\mean{cos(\theta)}$ vs. $z$ for in NB-FOPA ($D\simeq 0.02~\mathrm{ps}/\sqrt{\mathrm{km}}$).
Continuous curves, from top to bottom, refer to spun fibers with decreasing spin rate (10, 2, 0.5 turn/m); the dashed curve refer to unspun fibers. 
The corresponding mean gain are 26.5, 24.9, 22.9, 16.1 dB, respectively.}\label{fig:figura4}
\end{figure}

NPR effects become relevant for rapidly spun fibers, for which the averaging
of nonlinearity cannot be performed and the effective nonlinear coefficient is $\gamma$ \cite{mac06oe}. In fact,
as the spin rate increases, the peak in the STD, due to the NPR, shifts to
larger PMDC values because $D$ does not follow the same scaling law with $p$ of $L_C$ \cite{gal06oft}. Moreover, the mean gain of both BB- and NB-FOPA increases because of the enhancement in the mean pump-signal SOPs alignment 
$\mean{\hat p \cdot \hat s}=\mean{cos(\theta)}$, as already demonstrated in \cite{bet08ptl,gal08ptl}.
This effect is shown in Fig. \ref{fig:figura4} for the NB-FOPA.
Finally, observe that highly increasing the spin rate does not lead to a gain enhancement
similar to that observed for Raman and Brillouin amplifiers \cite{bet08ptl,gal08ptl}.
In fact, PMD affects the pump-signal alignment, restored by spinning,
but also the phase matching \cite{lin04ol,wil08jlt}, whose dependence on spinning is actually neglectable.

\section{Conclusions}\label{sec:concl}
In conclusion, the effects of polarization mode dispersion and unidirectional fiber spinning 
in broad-band and narrow-band single pump, fiber optical parametric amplifiers were
studied by means of numerical solutions of the governing equations.
The effects of nonlinear polarization rotation are relevant in 
such amplifiers, in particular when unidirectional spinning is applied.
The gain can be enhanced by spinning and, in NB-FOPA for
$10^{-2}\;\mathrm{ps}/\sqrt{\mathrm{km}}<D<4 \cdot 10^{-2}\; \mathrm{ps}/\sqrt{\mathrm{km}}$, the gain variance can be decreased.
This observation is relevant for slow and fast light generation, where
polarization mode dispersion highly affects the interaction \cite{wil08jlt,sch08jlt}.
The gain increase stems from the spinning enhancement of the alignment between the pump and the
signal along the fiber. 
However, differently from Raman and Brillouin amplifiers, the phase matching condition 
is also affected by polarization mode dispersion and the fiber spinning is ineffective 
to counteract mismatch. So, a limit to gain enhancement is found.
%

\end{document}